\documentclass{aa}
\usepackage{graphicx,lscape}
\usepackage{color}
\usepackage{txfonts,natbib}
\bibpunct{(}{)}{;}{a}{}{,}
\newcommand{\vsini} {$v$\,sin\,$i$}
\newcommand{\Teff} {T$_{\rm eff}$}
\newcommand{\grav} {log\,{\em g}}
\newcommand{\kms} {km\,s$^{-1}$}

\newcommand{\hii} {H\,{\sc ii}}

\begin{document}
   \title{The chemical composition of the Orion star-forming
   region\thanks{Based on observations made with the Nordic Optical
   Telescope, operated
   on the island of La Palma jointly by Denmark, Finland, Iceland,
   Norway, and Sweden, in the Spanish Observatorio del Roque de los
   Muchachos of the Instituto de Astrof\'isica de Canarias.}}
 
   \subtitle{III. C, N, Ne, Mg and Fe abundances in B-type stars revisited}

   \author{Mar\'ia-Fernanda Nieva\inst{1}\thanks{Visiting researcher at Dr. Karl Remeis-Sternwarte \& ECAP, 
          Universit\"at Erlangen-N\"urnberg.}
          \and
          Sergio Sim\'on-D\'iaz\inst{2,3}
          }

   \institute{Max-Planck-Institut f\"ur Astrophysik, Karl-Schwarzschild-Str. 1, D-85741
          Garching, Germany
         \and
         Instituto de Astrof\'isica de Canarias, E-38200 La Laguna, Tenerife,
         Spain
         \and
         Departamento de Astrof\'isica, Universidad de La Laguna, E-38205 La
         Laguna, Tenerife, Spain
             }

   \titlerunning{C, N, Ne, Mg and Fe abundances in Ori\,OB1 B-type stars}
   \authorrunning{Nieva \& Sim\'on-D\'{\i}az}
   \date{}

% \abstract{}{}{}{}{} 
% 5 {} token are mandatory
 
  \abstract
  % context heading (optional)
  % {} leave it empty if necessary  
  {
Early B-type stars are invaluable indicators for elemental
abundances of their birth environments. In contrast to the surrounding
neutral interstellar matter (ISM) and \hii\ regions their chemical 
composition is unaffected by depletion onto dust grains and by the 
derivation of different abundances from recombination and collisional lines. In combination with 
ISM or nebular gas-phase abundances they facilitate the otherwise
inaccessible dust-phase composition to be constrained.}
  % aims heading (mandatory)
   {
   Precise abundances of C, N, Mg, Ne, Fe in early B-type stars in the 
   Orion star-forming region are determined in order to: a)  review previous 
   determinations using a self-consistent quantitative spectral analysis based on 
   modern stellar atmospheres and recently updated model atoms, b) complement results 
   found in Paper~I for oxygen and silicon, c) establish an accurate and reliable set 
of stellar metal abundances to constrain the dust-phase composition of the Orion \ion{H}{ii} region in Paper~II of the series.}
  % methods heading (mandatory)
   {A detailed, self-consistent spectroscopic study of a sample of 13 narrow-lined 
   B0\,V-B2\,V stars in Ori\,OB1 is performed. High-quality spectra
   obtained with FIES@NOT are analysed using both a hybrid non-LTE method
   (classical line-blanketed LTE model atmospheres and non-LTE line
   formation) and line-profile fitting techniques, validating the approach 
   by comparison with results obtained in Paper~I using line-blanketed 
   non-LTE model atmospheres and a curve-of-growth analysis.}
  % results heading (mandatory)
   {The comparison of the two independent analysis strategies gives
   consistent results for basic stellar parameters and abundances of
   oxygen and silicon. The extended analysis to C, N, Mg, Ne, and Fe finds a high degree of 
   chemical homogeneity, with the 1$\sigma$-scatter typically adopting 
   values of 0.03--0.07 dex around the mean for the various elements.
   Present-day abundances from B-type stars in Ori\,OB1 are compatible at similar precision with
   cosmic abundance standard values as recently established from
   early-type stars in the solar neighbourhood and also with the Sun.    
}
  % conclusions heading (optional), leave it empty if necessary 
  {}


   \maketitle
%
%________________________________________________________________

\section{Introduction}\label{intro}

The Orion complex, containing the Orion molecular cloud and the 
Orion OB1 (Ori\,OB1) association, is one of the nearest Galactic regions
with ongoing star formation, located at a distance of about 400\,pc. 
It is unique for studying young massive stars, interstellar matter,
and their interaction via feedback processes. The Orion complex can be regarded 
as the best reference for any further study of other star-forming regions. Its proximity and 
brightness have facilitated many detailed investigations, that yielded
important discoveries and a wealth of information on many astrophysical 
phenomena over the past decades. 

\citet{b64} divided Ori\,OB1 into four subgroups of stars, namely Ia, Ib, Ic 
and Id, distinguished by their different location in the sky and by their ages. 
\citet{b94} derived mean ages of 11.4$\pm$1.9, 1.7$\pm$1.11, 4.6$\pm$2, and 
$<$1\,Myr for subgroups Ia to Id, respectively. The youngest subgroup 
Ori\,OB1~Id is associated with the Orion nebula (M\,42), the most studied 
\hii\ region and the closest ionized nebula to the Sun. 
The correlation between the ages of the stellar subgroups,
their location, and the large scale structures in the interstellar
medium around Orion\,OB1 have been interpreted as features of
sequential star formation and the impact of type-II supernovae
\citep{ro79,c79,b94}.

Early B-type stars in Ori\,OB1 are important astrophysical objects
that can be used to obtain present-day chemical abundances in terms of galactic chemical evolution.  
Moreover, the possibility of comparing the derived
stellar abundances with those resulting from the spectroscopic analysis
of the Orion nebula makes the tandem {\em massive stars in Ori\,OB1} and {\em M\,42} a 
keystone in the investigation of the reliability of B-type stars and \hii\ regions
as tracers of present-day abundances in the Universe.

First abundance determinations in B-type stars in Ori\,OB1 in the 90's 
\citep{gl92, cl92, cl94, k92, g98} indicated a large scatter 
up to 0.4\,--\,0.5\,dex for O and Si, but also $\ge$\,0.2
dex for other elements such as C, N, and Fe. At that time, the most extensive
work was carried out by \citet{cl92,cl94}, hereafter CL92 and CL94, respectively.
They derived C, N, O, Si, and Fe abundances of 18 B-type main sequence 
stars from the four subgroups comprising the Ori\,OB1 association. They
obtained a similar scatter in O and Si abundances as in the other cited works 
(based on smaller samples of stars), but also found that the highest abundances 
were associated to some of the stars in the youngest (Id and some Ic) 
subgroups\footnote{For other elements the scatter was also relatively large 
(e.g. 0.2, 0.45, and 0.35 dex for C, N, and Fe, respectively), but they did not 
find significant variations of abundances across the subgroups.}.
This abundance pattern was interpreted as an observational evidence supporting
a scenario of induced star formation in which the new generation of stars are formed 
from interstellar material contaminated by type-II supernovae (SN\,II) ejecta.

This picture has changed substantially after new studies of 
oxygen and silicon abundances of early B-type stars in Ori\,OB1 by 
\citet{sergio06} and \citet[hereafter Paper~I]{sergio10}. 
Both stellar parameters and chemical abundances were shown to have
different values than found previously, for the same stars. 
As a consequence, high chemical homogeneity of O and Si as derived
from early B-type stars the Orion star-forming region is implied, and 
no clear indication for contamination with SN\,II nucleosynthesis 
products is found. 

The results from Paper~I for the mean abundance and standard deviation for O and Si
 are in full agreement with recent findings 
of homogeneous present-day chemical abundances in the solar neighbourhood 
\citep[hereafter PNB08]{pnb08}. There, a small but 
representative sample of early B-type stars was analysed, obtaining
a significantly smaller scatter for  C, N, O,
Si, Mg, and Fe abundances than commonly found by previous studies.  

 Some reasons causing improvements in chemical abundance determinations
of B-type stars can be found in \citet{np10a,np10b}, and references therein. In brief, 
these are a consequence of the minimization of systematic uncertainties of observational 
origin and from the spectral modeling and spectral analysis, with two main
contributions: (a) the application of a {\em self-consistent} quantitative spectral analysis, 
i.e. to derive stellar parameters and chemical abundances from the spectrum 
only, without resorting to any extra ingredient like a photometric effective 
temperature estimation, and (b) the reliability of the 
atomic models and the selection of appropriate diagnostic lines for the stellar 
parameter and abundance determination.

In view of the results obtained in Paper I and PNB08 a revision
of abundances of other important elements in Ori\,OB1 stars 
is warranted. The main aim of this third paper of the series, in combination 
with Paper I, is to establish an accurate and reliable set of abundances for C, N, O, Ne, Mg,
Si, and Fe in early B-type stars in the Orion star-forming region, providing 
a basis to perform a thorough comparison between stellar and nebular chemical 
abundances in the Orion star-forming region \citet[hereafter Paper~II of this series]{sergioII10}.

The analysis presented in this paper is based on recently updated model atoms for 
non-LTE calculations (see Table~\ref{atoms}) which have been already applied in a 
small sample of early B-type stars in the solar neighbourhood (PNB08) and more 
recently in a larger sample  (Nieva \& Przybilla, in prep.). These model atoms 
are not available at present in the stellar atmosphere code used in Paper~I, 
therefore, we decided to base the analysis on the codes and techniques described in 
PNB08. For consistency, we performed the whole analysis from scratch (also including the 
stellar parameters, and O and Si abundance determination). The comparison of results
referent to stellar parameters, O and Si abundances obtained by two independent 
analyses will be very valuable to reinforce the conclusions presented in Paper I 
and/or investigate possible biases originated by the use of different stellar atmosphere codes and methodologies.
 
The paper is structured as follows: Sect.~\ref{observa} briefly describes the star sample 
and the observations. The spectrum synthesis in non-LTE and the codes used in the 
present work are described in Sect.~\ref{syn}. The self-consistent quantitative 
spectral analysis is summarized in Sect.~\ref{analysis}. Sect.~\ref{results} resumes 
the results and the comparison with other representative works in the literature. A 
summary is given and the conclusions are discussed in Sect.~\ref{summary}. 
An example of a global spectrum synthesis for one star of the sample is given in Appendix~\ref{appA}.

%__________________________________________________________________

\section{The star sample}\label{observa}

The star sample consists of 13 early B-type main sequence stars of spectral 
classes B0\,V to B2\,V and low projected rotational velocities 
(\vsini\,$\leq$\,60 \kms) in the Ori\,OB1 association. 
The spectroscopic dataset used in this study is the same
presented in Paper I. It refers to high resolution and 
high signal-to-noise ratio spectra obtained with the Fibre-fed Echelle 
Spectrograph (FIES) attached at the 2.5m Nordic Optical Telescope (NOT) in 
El Roque de los Muchachos observatory on La Palma (Canary Islands, Spain) 
in November, 2008. Details of the data reduction process and the quality
of the spectra can be found in Paper~I.

\section{Spectrum synthesis in non-LTE}\label{syn}

The non-LTE line-formation computations follow the methodology discussed in detail 
in our previous studies for H, He and C \citep{np07,np08} and for N, O, Ne, Mg, Si and Fe (PNB08).
In brief, a hybrid non-LTE approach is employed, i.e. combining
classical line-blanketed LTE model atmospheres and non-LTE
line-formation calculations. This technique provides an efficient way to 
compute realistic synthetic spectra in all cases where the atmos\-pheric 
structure is close to LTE, as it is the case for the early B-type main 
sequence stars analysed here \citep{np07}.

The model atmospheres were computed with the {\sc Atlas9}~code \citep{kur93b}
which assumes plane-parallel geometry, hydrostatic, radiative and local 
thermodynamic equilibrium, and chemical homogeneity. 
Solar abundances of \citet{gs98} were adopted in the atmospheric structure\footnote{The 
effect of the 'new' solar abundances (Asplund et al.~2009) on the atmospheric structure and on the ODF's is weak.} 
computations. Line blanketing was realized here by means of opacity distribution 
functions ODFs \citep[]{kur93a} with solar abundances scaled to \citet{gs98}.  
The model atmospheres were held fixed in the subsequent non-LTE calculations.

Non-LTE level populations and model spectra were obtained with
recent versions of {\sc Detail} and {\sc Surface} \citep{gid81,but_gid85}; 
both updated by K.\,Butler. The coupled radiative transfer and statistical
equilibrium equations were solved with {\sc Detail}, employing an accelerated
lambda itera\-tion scheme of \citet{rh91}. This allows even complex ions to be 
treated in a realistic and efficient way. Synthetic spectra were then calculated 
with {\sc Surface}, using refined line-broadening theories. The sources of model 
atoms adopted for the present work are listed in Table~\ref{atoms}. Details on 
the updates are given in by Nieva \& Przybilla (in prep.).

\begin{table}[t!]
\footnotesize
\caption[]{Model atoms for non-LTE calculations.\\[-3mm] \label{atoms}}%\\[2mm]
\setlength{\tabcolsep}{.15cm}
\begin{tabular}{ll}
\hline
\hline
\footnotesize
            Ion     &  Model atom \\
\hline\\[-3mm]
     H          &  \citet{pb04}\\
\ion{He}{i/ii}  &  \citet{p05}\\
\ion{C}{ii-iv}  &  \citet{np06,np08}\\
\ion{N}{ii}     &  \citet{pb01}$^*$\\
\ion{O}{i/ii}   &  \citet{p00}, \citet{bb88a}$^*$\\
\ion{Ne}{i/ii}  &  \citet{mb08}$^*$\\
\ion{Mg}{ii}    &  \citet{p01}\\
\ion{Si}{iii/iv}&  \citet{bb90a}\\
\ion{Fe}{ii/iii}&  \citet{b98}, \citet{m07}$^*$\\
            \hline\\[-5mm]
           \end{tabular}
\begin{list}{}{}
\item[$^*$]Updated by N. Przybilla (priv. comm.)
\end{list}
\end{table}

Continuous opacities due to hydrogen and helium were considered in non-LTE and metal 
line blocking was accounted for in LTE via Kurucz' ODFs.
Microturbulence was considered in a consistent way throughout all
computation steps: selection of appropriate ODFs for 
line blanketing and line blocking, atmospheric structure
calculations, determination of non-LTE populations 
and formal solution.

Grids of synthetic spectra for H/He, C, N, O, Mg and Si have been computed 
 in order to speed up the analysis. 
They cover effective temperatures from 15\,000 to 35\,000\,K in steps of 1000\,K, 
surface gravities from 3.0 to 4.5 in steps of 0.1\,dex, 
microturbulences from 0 to 8\,km\,s$^{-1}$ in steps of 2\,km\,s$^{-1}$ (for H/He, Si 
and O microturbulence reaches a value of 12\,km\,s$^{-1}$) and metal abundances within 1\,dex
centered on the Cosmic Abundance Standard proposed in PNB08 
in steps of 0.1\,dex. Hydrogen and helium abundances adopt PNB08 values.
The lower limit of the surface gravity for each value of 
temperature/microturbulence was constrained by the convergence of {\sc Atlas9}.
All grids have been successfully tested by reproducing results from PNB08 within a broad temperature range.
Micro-grids varying abundances only,  were computed per star for neon and iron once all stellar 
parameters have been determined with the larger pre-computed grids.

\section{Self-consistent quantitative spectral analysis}\label{analysis}

The analysis method used in this paper is based on a spectral line-fitting 
procedure\footnote{Note that we used here a different approach than in Paper I. More
details can be found in Sect.~\ref{sec61}.}.
We aim at deriving atmospheric parameters and chemical abundances self-consistently
by reproducing several independent spectroscopic indicators simultaneously.
The atmospheric parameters primarily derived here are 
the effective temperature $T_\mathrm{eff}$, surface gravity $\log g$, 
microturbulence $\xi$, macroturbulence $\zeta$, projected
rotational velocity $v\,\sin\,i$, elemental abundances $\varepsilon$(X).
 
The quantitative analysis is mostly automatized, but it also allows to work 
interactively in some decisive steps. The interpolation in stellar parameters 
via spectral line fitting using $\chi^2$-minimization  within the precomputed 
grids is performed with the program {\sc Spas}\footnote {Spectrum Plotting and 
Analysing Suite \citep[{\sc Spas},][]{hirsch09}.}.  
The code allows flexible selection of spectroscopic indicators  
for parameter determinations which may vary from star to star 
upon availability of specific spectral lines.
Linelists can also be chosen interactively. This is crucial since then problematic spectral lines, 
i.e. too weak, blended, with low S/N, in a region where the continuum normalization is difficult or with a particular shortcoming in the spectral synthesis can be excluded.
The local continuum can also be adjusted interactively. This might 
slow down the analysis process in cases where the linelist is long but the gain in precision is high.

\begin{table}[t!]
\centering
\caption[]{
Spectroscopic indicators for the $T_\mathrm{eff}$ and $\log g$ determination. The boxes indicate 
ionization equilibria that have been quantitatively matched in the analysis.\\[-3mm]\label{indicators}}
 \setlength{\tabcolsep}{.06cm}
 \tiny
 \begin{tabular}{llcc@{\hspace{.1mm}}lc@{\hspace{.1mm}}c@{\hspace{.1mm}}cccc@{\hspace{.1mm}}cc@{\hspace{.1mm}}cc@{\hspace{.1mm}}c}
 \noalign{}
\hline
\hline
\#&Object& H & \ion{He}{i} & \ion{He}{ii} & \ion{C}{ii} & \ion{C}{iii}  & \ion{C}{iv} & \ion{O}{i} & \ion{O}{ii} & \ion{Ne}{i} &  \ion{Ne}{ii} & \ion{Si}{iii}& \ion{Si}{iv} & \ion{Fe}{ii}& \ion{Fe}{iii}\\
\hline
1 & \object{HD36512}  & $\bullet$ & \multicolumn{2}{c}{\fbox{$\bullet$~~~~~$\bullet$}} &\multicolumn{3}{c}{\fbox{\,$\bullet$~~~~~$\bullet$~~~~~$\bullet$\,}}  &&$\bullet$                                         &&  $\bullet$                                        & \multicolumn{2}{c}{\fbox{$\bullet$~~~~~$\bullet$}}&& $\bullet$\\
2 & \object{HD37020}  & $\bullet$ & \multicolumn{2}{c}{\fbox{$\bullet$~~~~~$\bullet$}} & \multicolumn{2}{c}{\fbox{$\bullet$~~~~~$\bullet$}}     &             && $\bullet$                                         & \multicolumn{2}{c}{\fbox{$\bullet$~~~~~$\bullet$}}& \multicolumn{2}{c}{\fbox{$\bullet$~~~~~$\bullet$}}&& $\bullet$\\
3 & \object{HD36960}  & $\bullet$ & \multicolumn{2}{c}{\fbox{$\bullet$~~~~~$\bullet$}} & \multicolumn{2}{c}{\fbox{$\bullet$~~~~~$\bullet$}}     &             && $\bullet$                                         & \multicolumn{2}{c}{\fbox{$\bullet$~~~~~$\bullet$}}& \multicolumn{2}{c}{\fbox{$\bullet$~~~~~$\bullet$}}&& $\bullet$\\
4 & \object{HD37042}  & $\bullet$ & \multicolumn{2}{c}{\fbox{$\bullet$~~~~~$\bullet$}} & \multicolumn{2}{c}{\fbox{$\bullet$~~~~~$\bullet$}}     &             && $\bullet$                                         & \multicolumn{2}{c}{\fbox{$\bullet$~~~~~$\bullet$}}& \multicolumn{2}{c}{\fbox{$\bullet$~~~~~$\bullet$}}&& $\bullet$\\
5 & \object{HD36591}  & $\bullet$ & \multicolumn{2}{c}{\fbox{$\bullet$~~~~~$\bullet$}} & \multicolumn{2}{c}{\fbox{$\bullet$~~~~~$\bullet$}}     &             &  \multicolumn{2}{c}{\fbox{$\bullet$~~~~~$\bullet$}}& \multicolumn{2}{c}{\fbox{$\bullet$~~~~~$\bullet$}}& \multicolumn{2}{c}{\fbox{$\bullet$~~~~~$\bullet$}}&& $\bullet$\\     
6 & \object{HD36959}  & $\bullet$ & \multicolumn{2}{c}{\fbox{$\bullet$~~~~~$\bullet$}} & \multicolumn{2}{c}{\fbox{$\bullet$~~~~~$\bullet$}}     &             &  \multicolumn{2}{c}{\fbox{$\bullet$~~~~~$\bullet$}}& \multicolumn{2}{c}{\fbox{$\bullet$~~~~~$\bullet$}}& \multicolumn{2}{c}{\fbox{$\bullet$~~~~~$\bullet$}}&& $\bullet$\\
7 & \object{HD37744}  & $\bullet$ & $\bullet$  &                                       & \multicolumn{2}{c}{\fbox{$\bullet$~~~~~$\bullet$}}     &              & \multicolumn{2}{c}{\fbox{$\bullet$~~~~~$\bullet$}}& \multicolumn{2}{c}{\fbox{$\bullet$~~~~~$\bullet$}}& \multicolumn{2}{c}{\fbox{$\bullet$~~~~~$\bullet$}}&\multicolumn{2}{c}{\fbox{$\bullet$~~~~~$\bullet$}}\\
8 & \object{HD35299}  & $\bullet$ & $\bullet$  &                                       & \multicolumn{2}{c}{\fbox{$\bullet$~~~~~$\bullet$}}     &              & \multicolumn{2}{c}{\fbox{$\bullet$~~~~~$\bullet$}}& $\bullet$ && \multicolumn{2}{c}{\fbox{$\bullet$~~~~~$\bullet$}}&\multicolumn{2}{c}{\fbox{$\bullet$~~~~~$\bullet$}}\\
9 & \object{HD36285}  & $\bullet$ & $\bullet$  &                                       & \multicolumn{2}{c}{\fbox{$\bullet$~~~~~$\bullet$}}     &              & \multicolumn{2}{c}{\fbox{$\bullet$~~~~~$\bullet$}}& $\bullet$ && \multicolumn{2}{c}{\fbox{$\bullet$~~~~~$\bullet$}}&\multicolumn{2}{c}{\fbox{$\bullet$~~~~~$\bullet$}}\\
10& \object{HD35039}  & $\bullet$ & $\bullet$  &                                       & \multicolumn{2}{c}{\fbox{$\bullet$~~~~~$\bullet$}}     &              & \multicolumn{2}{c}{\fbox{$\bullet$~~~~~$\bullet$}}& $\bullet$ && \multicolumn{2}{c}{\fbox{$\bullet$~~~~~$\bullet$}}&\multicolumn{2}{c}{\fbox{$\bullet$~~~~~$\bullet$}}\\
11& \object{HD36629}  & $\bullet$ & $\bullet$  &                                       & \multicolumn{2}{c}{\fbox{$\bullet$~~~~~$\bullet$}}     &              & \multicolumn{2}{c}{\fbox{$\bullet$~~~~~$\bullet$}}& $\bullet$ &&$\bullet$ &&\multicolumn{2}{c}{\fbox{$\bullet$~~~~~$\bullet$}}\\
12& \object{HD36430}  & $\bullet$ & $\bullet$  &                                       & \multicolumn{2}{c}{\fbox{$\bullet$~~~~~$\bullet$}}     &              & \multicolumn{2}{c}{\fbox{$\bullet$~~~~~$\bullet$}}& $\bullet$ &&$\bullet$ &&\multicolumn{2}{c}{\fbox{$\bullet$~~~~~$\bullet$}}\\
13& \object{HD35912}  & $\bullet$ & $\bullet$  &                                       & \multicolumn{2}{c}{\fbox{$\bullet$~~~~~$\bullet$}}     &              & \multicolumn{2}{c}{\fbox{$\bullet$~~~~~$\bullet$}}& $\bullet$ &&$\bullet$ &&\multicolumn{2}{c}{\fbox{$\bullet$~~~~~$\bullet$}}\\[1mm]
\hline
\end{tabular}
\end{table}

\begin{table*}
\setlength{\tabcolsep}{.08cm}
\caption[]{Stellar parameters and elemental abundances of the program stars. 
\\[-6mm]\label{parameters}}
%\small
\begin{tabular}{llccccr@{\,$\pm$\,}lr@{\,$\pm$\,}lr@{\,$\pm$\,}lr@{\,$\pm$\,}lr@{\,$\pm$\,}lcr@{\,$\pm$\,}lr@{\,$\pm$\,}lr@{\,$\pm$\,}lr@{\,$\pm$\,}lr@{\,$\pm$\,}lr}
\hline
\hline
& & \multicolumn{1}{c}{\raisebox{-0.65ex}{Sub-}} &
\multicolumn{2}{c}{\raisebox{-0.65ex}{FW}} & & 
\multicolumn{10}{c}{\raisebox{-0.65ex}{ADS}} & &
\multicolumn{10}{c}{\raisebox{-0.65ex}{$\log$\,X/H\,$+$\,12 (NLTE, ADS)}}\\
\cline{4-5} \cline{7-16} \cline{18-27}
\multicolumn{1}{c}{\raisebox{-0.65ex}{Target}}& 
\multicolumn{1}{c}{\raisebox{-0.65ex}{SpT}}   & 
\multicolumn{1}{c}{\raisebox{-0.65ex}{Gr.}} &
\multicolumn{1}{c}{\raisebox{-0.65ex}{$T_\mathrm{eff}$}} & 
\multicolumn{1}{c}{\raisebox{-0.65ex}{$\log g$}} & & 
\multicolumn{2}{c}{\raisebox{-0.65ex}{$T_\mathrm{eff}$}} &
\multicolumn{2}{c}{\raisebox{-0.65ex}{$\log g$}} & 
\multicolumn{2}{c}{\raisebox{-0.65ex}{$\xi$}} &
\multicolumn{2}{c}{\raisebox{-0.65ex}{v$\sin$i}} &
\multicolumn{2}{c}{\raisebox{-0.65ex}{$\zeta$}} & &  
\multicolumn{2}{c}{\raisebox{-0.65ex}{C}} & 
\multicolumn{2}{c}{\raisebox{-0.65ex}{N}} & 
\multicolumn{2}{c}{\raisebox{-0.65ex}{Ne}}&
\multicolumn{2}{c}{\raisebox{-0.65ex}{Mg}}&
\multicolumn{2}{c}{\raisebox{-0.65ex}{Fe}}
& \#\\
\cline{11-16}
& & & \multicolumn{1}{c}{\raisebox{-0.65ex}{(K)}} &
\multicolumn{1}{c}{\raisebox{-0.65ex}{(cgs)}} & &
\multicolumn{2}{c}{\raisebox{-0.65ex}{(K)}} &
\multicolumn{2}{c}{\raisebox{-0.65ex}{(cgs)}} &
\multicolumn{6}{c}{\raisebox{-0.65ex}{(km\,s$^{-1}$)}}\\
\hline                                                                              
                                                                                    
       %O
\object{36512} & B0\,V   &  Ic & 33700 & 4.2 & & 33400 & 300 & 4.30 & 0.07 & 4 & 1 &20 &  1 & 10 & 2 & & 8.36 & 0.15 & 7.83 & 0.15 & 8.16 & 0.10 &\multicolumn{2}{c}{7.51}  & 7.55 & 0.02 &  1\\ 
\object{37020} & B0.5\,V &  Id & 30500 & 4.2 & & 30700 & 300 & 4.30 & 0.08 & 0 & 1 &45 &  3 & 20 & 5 & & 8.40 & 0.07 & 7.85 & 0.09 & 8.09 & 0.08 &\multicolumn{2}{c}{7.64}  & 7.53 & 0.08 &  2\\ 
\object{36960} & B0.5\,V &  Ic & 28900 & 3.9 & & 29000 & 350 & 4.10 & 0.08 & 4 & 1 &28 &  2 & 20 & 6 & & 8.39 & 0.07 & 7.78 & 0.08 & 8.13 & 0.09 &\multicolumn{2}{c}{7.61}  & 7.53 & 0.11 &  3\\ 
\object{37042} & B0.7\,V &  Id & 29700 & 4.2 & & 29300 & 300 & 4.30 & 0.09 & 2 & 1 &30 &  2 & 10 & 3 & & 8.33 & 0.11 & 8.04 & 0.08 & 8.13 & 0.09 &\multicolumn{2}{c}{7.64}  & 7.54 & 0.09 &  4\\ 
\object{36591} & B1\,V   &  Ib & 27200 & 4.1 & & 27000 & 250 & 4.12 & 0.05 & 2 & 1 &12 &  1 & \multicolumn{2}{c}{\ldots}& & 8.30 & 0.10 & 7.81 & 0.09 & 8.10 & 0.07 & 7.61 & 0.02 & 7.49 &0.12 &  5\\         
\object{36959} & B1\,V   &  Ic & 25900 & 4.2 & & 26100 & 200 & 4.25 & 0.07 & 0 & 1 &12 &  1 &  5 & 1 & & 8.37 & 0.11 & 7.85 & 0.09 & 8.18 & 0.10 & 7.66 & 0.04 & 7.53 &0.11 &  6\\               
\object{37744} & B1.5\,V &  Ib & 23800 & 4.1 & & 24000 & 400 & 4.10 & 0.10 & 0 & 1 &33 &  2 & 15 & 2 & & 8.32 & 0.07 & 7.81 & 0.10 & 8.12 & 0.08 &\multicolumn{2}{c}{7.58}  & 7.52 & 0.07 &  7\\ 
\object{35299} & B1.5\,V &  Ia & 23700 & 4.2 & & 24000 & 200 & 4.20 & 0.08 & 0 & 1 & 8 &  1 & \multicolumn{2}{c}{\ldots}& & 8.37 & 0.07 & 7.81 & 0.07 & 8.12 & 0.09 & 7.61 & 0.07 & 7.49 &0.10 &  8\\               
\object{36285} & B2\,V   &  Ic & 20600 & 4.0 & & 21700 & 300 & 4.25 & 0.08 & 0 & 1 &11 &  1 &  8 & 1 & & 8.32 & 0.07 & 7.77 & 0.09 & 8.04 & 0.08 & 7.52 & 0.08 & 7.54 &0.10 &  9\\               
\object{35039} & B2\,V   &  Ia & 19800 & 3.7 & & 19600 & 200 & 3.56 & 0.07 & 4 & 1 &12 &  1 &  7 & 1 & & 8.34 & 0.10 & 7.77 & 0.09 & 8.04 & 0.08 & 7.49 & 0.06 & 7.48 &0.10 & 10\\              
\object{36629} & B2\,V   &  Ic & 20000 & 4.1 & & 20300 & 400 & 4.15 & 0.10 & 2 & 1 &10 &  1 &  5 & 1 & & 8.36 & 0.07 & 7.82 & 0.08 & 8.02 & 0.09 & 7.50 & 0.02 & 7.52 &0.11 & 11\\               
\object{36430} & B2\,V   &  Ic & 18600 & 4.1 & & 19300 & 200 & 4.14 & 0.05 & 0 & 1 &20 &  2 & 10 & 2 & & 8.32 & 0.08 & 7.76 & 0.10 & 8.03 & 0.12 & 7.54 & 0.05 & 7.50 &0.07 & 12\\               
\object{35912} & B2\,V   &  Ia & 18500 & 4.0 & & 19000 & 300 & 4.00 & 0.10 & 2 & 1 &15 &  1 &  8 & 1 & & 8.33 & 0.09 & 7.76 & 0.07 & 8.05 & 0.11 & 7.50 & 0.05 & 7.52 &0.08 & 13\\               
\hline
\end{tabular}
\tablefoot{Target id are HD numbers. Parameters derived with {\sc Fastwind} 
(FW, Paper I) and {\sc Atlas+Detail+Surface} (ADS, this work)
are listed. Uncertainties in parameters from Paper~I are similar than in the
present work.}
\end{table*}

\begin{table*}[t!]
\centering
\caption[]{Average metal abundances.}.\\[-6mm]\label{abundances}%\\[2mm]
 \begin{tabular}{ccccccc}
 \noalign{}
\hline
\hline
\\[-2mm]
 & This work & Paper~I & PNB08 & CL94$^\mathrm{NLTE}$ & CL94$^\mathrm{LTE}$ & Sun\tablefootmark{a} \\ %$T_\mathrm{eff}$,
\hline
\\[-2mm]
C  & 8.35$\pm$0.03  [0.09]  &                      & 8.32$\pm$0.03 & 8.40$\pm$0.11 & 8.35$\pm$0.06$~$     & 8.47$\pm$0.05 \\
N  & 7.82$\pm$0.07  [0.09]  &                      & 7.76$\pm$0.05 & 7.76$\pm$0.13 & 7.79$\pm$0.11$~$     & 7.87$\pm$0.05 \\
O  & 8.77$\pm$0.03  [0.10]  & 8.73$\pm$0.04        & 8.76$\pm$0.03 & 8.72$\pm$0.13 & 8.73$\pm$0.13$~$     & 8.73$\pm$0.05 \\ 
Ne & 8.09$\pm$0.05  [0.09]  &                      & 8.08$\pm$0.03 & 8.11$\pm$0.04\tablefootmark{b}&$\dots$    & 7.97$\pm$0.10 \\
Mg & 7.57$\pm$0.06  [0.05]  &                      & 7.56$\pm$0.05 & $\dots$       & $\dots$              & 7.64$\pm$0.04 \\
Si & 7.50$\pm$0.06  [0.06]  & 7.51$\pm$0.03        & 7.50$\pm$0.02 & 7.14$\pm$0.13 & 7.36$\pm$0.14$~$     & 7.55$\pm$0.03 \\
Fe & 7.52$\pm$0.02  [0.09]  &                      & 7.44$\pm$0.04 & $\dots$       & 7.47$\pm$0.10$~$     & 7.54$\pm$0.04 \\
\hline\\[-2mm] 
\end{tabular}
\tablefoot{Average metal abundances, 1-$\sigma$ uncertainties 
and intrinsic uncertainties for each element (in brackets) 
derived in this work are compared to values from Paper~I, 
the present-day Cosmic Abundance Standard by PNB08, 
abundances in NLTE and LTE of 17 stars in Ori\,OB1 by CL94, 
and solar abundances. 
\tablefoottext{a}{Protosolar values by AGSS09 (Table 5 in the original paper).}
\tablefoottext{b}{Value from 16 stars, from CHL06.}
}
\end{table*}

The strategy used for the analysis of our star sample is as 
follows. A first estimation of $T_\mathrm{eff}$ and $\log g$ is obtained by means of a 
simultaneous fit to most H and He lines (an accuracy better than $\sim$\,5\% and 
0.1--0.2\,dex, respectively, is expected for the case of this high quality set 
stellar spectra). Then the procedure starts to consider lines of other elements 
 to end up with a self-consistent solution for atmospheric parameters 
and chemical abundances, also including an estimation of their internal errors.
In this second step, every element is analysed independently and some interactive iterations 
to fine-tune the stellar parameter determination are needed in order to find a unique 
solution for all indicators.
 
Numerous metal lines were analysed simultaneously per star/element. 
The high S/N and high resolution spectra, the optimum continuum normalization,
the ample wavelength range coverage, the presence of many unblended sharp spectral lines and the adoption of our updated model atoms for spectral synthesis calculations
allow us to analyse more lines than in standard works.
The linelist varies slightly from star to star because the selection of 
good lines to be analysed depends on the temperature, the quality of the spectra
and the wavelength coverage. In particular, the line lists used to derive 
final values of elemental abundances has been chosen for each star, in a way that 
the abundance from any line does not differ by more than $\sim$2--3$\sigma$ from average.

Ionization equilibria, i.e.~the requirement that lines from different ions of 
an element have to indicate the same chemical abundance, along with the fitting of
the wings of the H Balmer lines facilitate a fine-tuning of the previously derived parameters.
Whenever possible, we use multiple ionization equilibria for the 
parameter determination as the redundancy of information helps to minimize 
systematic errors. In particular, we have found that the use of more than one
ionization equilibrium helps to better constraint the surface gravity compared to
the case when only H lines are used.

The selection of ionization equilibria to be analysed depends on the temperature
range of the star. Table~\ref{indicators} summarizes the available
indicators\footnote{Note that we analyse Si\,{\sc iii/iv} lines only because 
the model atom employed here does not produce reliable Si\,{\sc ii} lines (see also 
Paper~I).} for the star sample. Note that the use of more than one 
ionization equilibrium has been always possible. In addition, the availability of 
the Fe\,{\sc ii/iii} ionization equilibrium allow to also impose a constraint to 
the metallicity, [M/H]\,$\simeq$\,[Fe/H]. 

The fine-tuning of microturbulence, stellar parameters and abundances 
is performed at the same time. In particular, microturbulence is determined by imposing 
a zero correlation between individual line abundances for some ions.

The strength of this analysis method is that most important sources of systematic
uncertainties can be identified and the remaining errors reduced to minimum
values. See~\citet{np10a, np10b} for more detailed discussions on the methodology
and systematic errors in the spectral analysis of this kind of stars.

\section{Our results for the Orion B-type stars 
and comparison with previous work}\label{results}

Table~\ref{parameters} summarizes atmospheric parameters 
and elemental abundances of C, N, Ne, Mg and Fe  derived in 
this study for the 13 sample stars. There, also spectral type and 
subgroup to which the stars belong within Ori\,OB1 are given.
The identification numbering of the stars (last column) follows 
the one used in Paper~I.
The corresponding mean values and 1-$\sigma$ uncertainties derived 
from the whole star sample are presented in first column of 
Table~\ref{abundances}. Note that mean abundances for O and Si obtained in 
this work are also indicated. These abundances (though not presented in 
Table \ref{parameters}) were derived for completeness, and to be compared 
with those obtained in Paper I (see Sect.~\ref{sec61}).

Table~\ref{abundances}
also contains
information about results obtained by previous studies of early B-type stars
in Ori\,OB1 from Paper~I, CL92, CL94 and \citet{c06}, hereafter CHL06, early B-type stars in the solar neighbourhood from PNB08, 
and the set of protosolar abundances proposed by \citet{a09}, hereafter AGSS09. 
Values from CL94 in Table~\ref{abundances}  have been computed from their total sample and are 
slightly different than those values indicated by them, calculated from a sub-sample of their work.
A detailed 
discussion on the comparison of results from the analysis performed here and the various
cited studies is presented in the following sub-sections.

\begin{figure*}[ht!]
\centering
\vspace{1cm}
\resizebox{0.95\hsize}{!}{\includegraphics{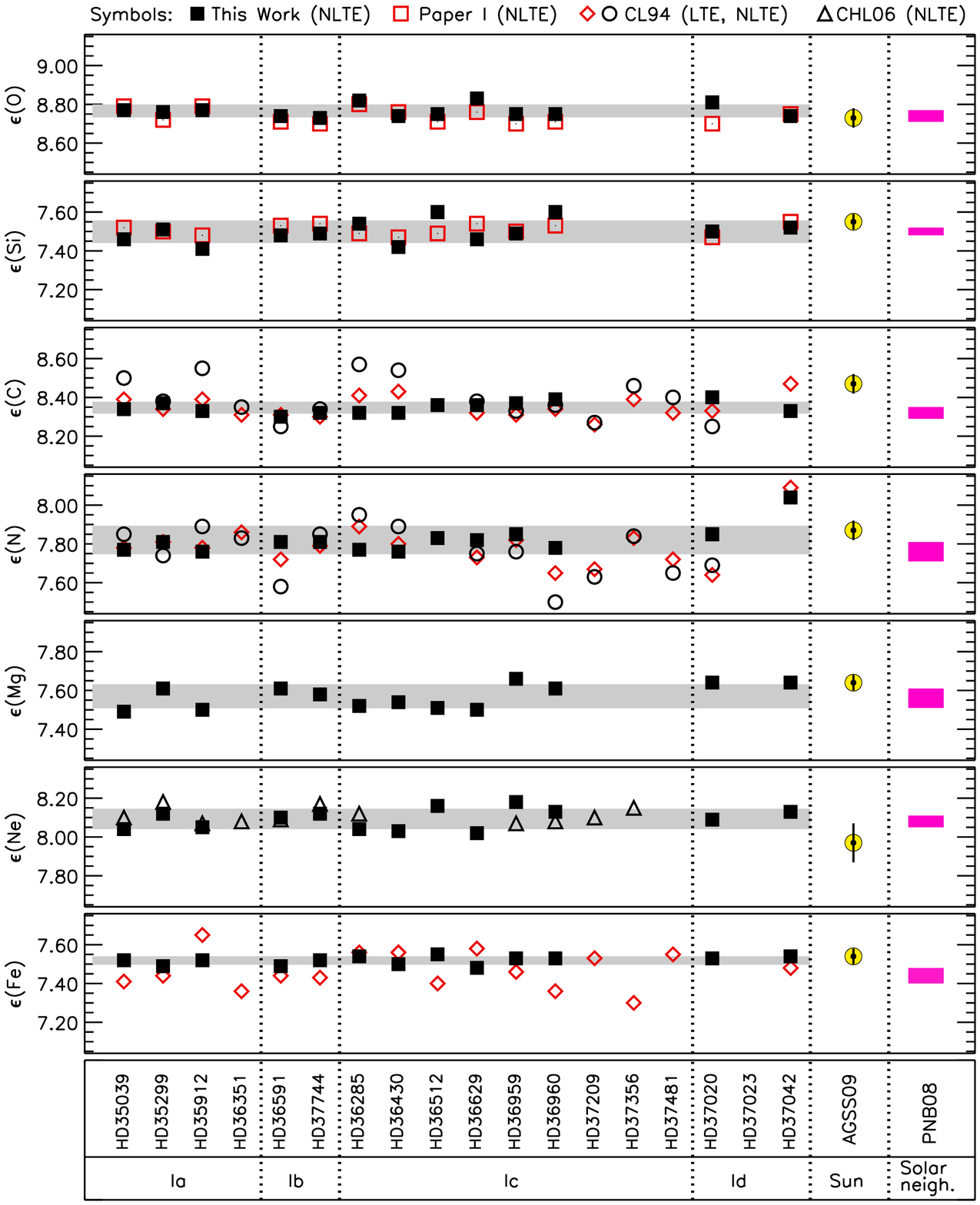}}\\
\caption[]{
Elemental abundances in our Ori OB1 sample stars and a comparison with
previous work, and with common abundance standards. 
Symbols for data from this work, Paper I, CL94 and CHL06 are explained in 
the upper part of the figure. The grey-shaded region represents the
1-$\sigma$ dispersion of abundances from the whole sample of 13 stars. Protosolar
values of AGSS09 and the Cosmic Abundance Standard of PNB08 are given on the right hand side of the figure for comparison.}
\label{Abund_all}
\end{figure*}

>From inspection of Tables~\ref{parameters} and \ref{abundances} it can be 
concluded that the large scatter in abundances found for C, N, and Fe by 
previous works (see Sect. \ref{intro}) has been significantly reduced. In
addition, the high degree of homogeneity of derived abundances 
for all the analysed elements is remarkable (the 1-$\sigma$
dispersion associated to the mean value of the whole sample is always
significantly smaller than the intrinsic uncertainties from the individual
analysis,  obtained by averaging the individual errors for each star listed in Table \ref{parameters}. These values are indicated in brackets in Table~\ref{abundances}). 
To better illustrate these conclusion, we plot in Fig.~\ref{Abund_all},
the individual abundances for the analysed Ori\,OB1 stars (black filled squares) 
together with the 1-$\sigma$ dispersion of abundances derived from the whole
sample (grey-shaded area), and compare them with other studies listed in Table~\ref{abundances}.

\begin{figure*}[t!]
\centering
\resizebox{0.8\hsize}{!}{\includegraphics[angle=90]{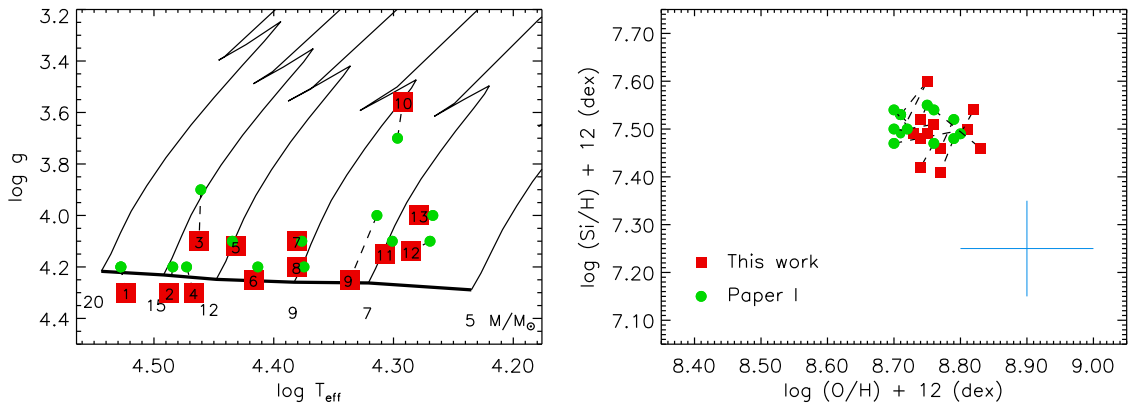}}\\
\caption[]{Comparison of effective temperature and surface
gravities (left panel) and oxygen and silicon abundances (right panel) derived in 
Paper~I with {\sc Fastwind} and the present work with {\sc Atlas+Detail+Surface}. 
Evolutionary tracks (corresponding to Z=0.02) are extracted from \citet{Sch92}. Numbers in
the left panel identify the stars, as in Table~\ref{parameters} 
The blue cross in the right panel indicates the typical uncertainties in O and Si 
abundances from individual analyses.} 
\label{ADSparametersFW}
\end{figure*}

\subsection{B stars in Ori\,OB1: Paper~I}\label{sec61}

 The same set of observed spectra was analysed in Paper~I and in the present work.
The method used here based on {\sc Atlas+Detail+Surface} and the one adopted in 
Paper~I ({\sc Fastwind}) employ the same philosophy:  the simultaneous determination 
of stellar parameters and chemical abundances from the spectrum only. However, the 
codes and the spectral analysis are different. Results of Paper~I are based on the 
unified atmosphere code {\sc Fastwind} \citep{pu05} that models the stellar atmosphere 
and wind in non-LTE. The input atomic data for the line formation of hydrogen, helium, 
silicon and oxygen are not identical to the ones used here. Furthermore, the 
determination of chemical abundances in Paper~I was based on equivalent widths and the 
curves of growth, while here direct spectral line fits are performed.  Finally,
the spectroscopic diagnostics used for the stellar parameters determination are 
different: on the one hand, Paper~I relies exclusively on the H Balmer lines, along 
with the \ion{He}{i/ii} and/or \ion{Si}{ii/iii/iv} ionization equilibria;
on the other hand, as described in Sect.\ref{analysis}, the analysis presented in this
paper uses all available ionization equilibria for He, Si, C, N, O, Ne, and Fe.

Stellar parameter and oxygen and silicon abundance determinations were carried out 
independently to the work in Paper~I, without a priori information of final results. 
The comparison of both studies can be  found in 
Tables~\ref{parameters} (stellar parameters) and \ref{abundances} (mean 
abundances and 1-$\sigma$
dispersion), and Figs.~\ref{ADSparametersFW} and \ref{Abund_all}. 
In spite of all differences in the two analyses, the similarity of results derived
in both works is remarkable. The global agreement in the four quantities, \Teff, 
\grav, $\epsilon$(O), and $\epsilon$(Si), is very good within the uncertainties.

Fig.~\ref{Abund_all} shows a detailed comparison of abundances for the 13 sample stars 
with results for O and Si from Paper~I (red open squares) for the same objects.  
The analysed stars are separated in the various Ori\,OB1 stellar subgroups.
No trend in O or Si is found, confirming the conclusions of Paper~I of chemical homogeneity 
of these elements in the region. We refer the reader to Paper~I for a similar comparison
with results from CL92 and CL94.
The differences of mean abundances for the sample is 0.04\,dex for oxygen and 0.01\,dex 
for silicon, that within the errors are negligible, and the 1-$\sigma$ dispersion of 
abundances in the whole sample obtained in both analysis is similar (see Table~\ref{abundances}).

\subsection{B stars in Ori\,OB1: Cunha et al.}

The chemical composition of the Orion star forming region derived by CL94 has been a 
commonly used reference for several applications in contemporary astrophysics for the past two decades.
 They derived LTE and non-LTE abundances for a sample of 18 B-type stars. Both
analyses were based on a photometric determination of the effective temperature. LTE abundances were obtained
using ATLAS6 line blanketed model atmospheres \citep{Kur79}. Non-LTE abundances were derived using
a similar strategy as in the present work, as described in Sect. \ref{analysis}; i.e {\sc Detail} and 
{\sc Surface} NLTE computations on top of a line blanketed LTE atmospheric structure. Details on the non-LTE computations used by CL94 can be found 
in the original papers by \citet{bb88b,bb89, bb90a,bb90b} and \citet{eb88}. 
Two main differences distinguish
our NLTE computations from those used by CL94: on the one hand, older versions of the codes and mostly
older atomic data (our model atoms are listed in Table~\ref{atoms}), except for the Si model atom that 
is the same for both works; and on the other hand, their non-LTE line formation was computed on top of the slightly blanketed 
LTE \citet{g84} atmospheric structures, while we use the fully blanketed ATLAS9 model.

 In CL94, the idea of self-enrichment via supernovae explosions was very attractive to 
explain discrepant abundances between star members of the region. 
This explanation to the chemical patterns found by CL92 and CL94 has been challenged by new 
results from \citet{sergio06} and Paper~I. There, highly homogeneous O and Si abundance was 
found for Ori\,OB1, with no clear indications of SN-II contamination. 
The high degree of homogeneity is confirmed here by means of an independent spectroscopic analysis. 

A comparison of C, N, and Fe abundances derived from our analysis, 
with those provided by CL94  (both NLTE\footnote{ The NLTE abundances have been
the preferred cited values in the literature} and
LTE) is presented in Table~\ref{abundances} (mean abundances and 1-$\sigma$ dispersions) and Fig.~\ref{Abund_all} 
 (star-by-star abundances). 
 Despite the differences in the methods, codes and observed spectra, the mean values of the present work 
and those from CL94 for C, N, O, and Fe, both in non-LTE and in LTE, are in agreement within the uncertainties. Our chemical 
abundances are, however, more precise thanks to the improvements in reducing several systematic effects in 
the whole procedure \citep{np10a,np10b}. In particular, Fig.~\ref{Abund_all} shows significant 
differences in the derived abundances for individual stars, e.g. of up to 0.3 dex in some stars for the case of C and N
from CL94, in contrast to our results. It is also remarkable the case of iron, where we find a certainly 
smaller scatter compared to their LTE determinations.
Only one element has drastically changed its mean abundance: silicon.
The discrepancy of Si abundance is probably related to the combined effect of an inaccurate 
determination of the effective temperature and the use of certain diagnostic lines, traditionally 
used to establish the microturbulence and derive the Si abundance, which have been identified
to be problematic (see Paper~I for more details).

In a more recent work by CHL06, non-LTE Ne abundances for a sub-sample of stars in CL94 
were derived using the previously determined stellar parameters and synthetic spectra 
computed with {\sc Tlusty} \citep{h88,hl95}. In brief, the code calculates the stellar structure and the line formation simultaneously in NLTE. The non-LTE atmospheric structures from {\sc Tlusty} are in agreement to better than 1$\%$ with those from ATLAS9, in LTE, for temperatures up to (at least) $\sim$35,000 K (see Nieva \& Przybilla 2007). Therefore the main difference to our work is the input atomic data
for the line formation. Moreover, their abundances were derived from \ion{Ne}{i}, while ours were determined from \ion{Ne}{i} and \ion{Ne}{ii}, that allowed us to analyse the hotter objects.
A good agreement  (star-by-star, and therefore mean abundances and scatter) is obtained between our Ne abundances and 
those derived in CHL06.

\subsection{B-type stars in the Solar neighbourhood}\label{standard}

The chemical abundances of the star sample agree very well with the 
recently proposed Present-Day Cosmic Abundance Standard of the solar neighbourhood,
obtained from a representative group of early B-type stars in associations and the field 
by PNB08 (see Table~\ref{abundances} and Fig.~\ref{Abund_all}). This agreement 
strengthens the idea that chemical homogeneity is naturally found in the solar neighbourhood 
once most prominent systematic uncertainties of stellar parameters and abundances have been minimized. 
These results are also confirmed by the study of a larger sample of B-type stars in the 
solar neighbourhood (Nieva \& Przybilla, in prep.). 

\subsection{The Sun}

Table~\ref{abundances} and Fig.~\ref{Abund_all} show the
overall agreement between present-day metal abundances in the Orion star-forming region 
as derived here from early B-type stars and solar\footnote{Note that protosolar abundances have been chosen because they represent the chemical composition at the Sun-formation time.} values by 
\citet{a09}, despite some small remaining discrepancies. 

Comparisons of chemical composition of young early B-type stars and a much older star like 
our Sun have a historical character, since the Sun is the preferred reference for the 
chemical abundances in most astrophysical applications. There are, however, 
several issues related to the Sun in context of Galactochemical
evolution that are still not well understood, e.g. whether the Sun has
actually migrated from its nursery to the current location in the
Galaxy \citep{port09,s09}, or some small-scale peculiarities in the solar 
abundance pattern in comparison with solar analogs \citep{jorge09,ivan09}.
The metal enrichment of the interstellar matter since the
Sun's formation is also not precisely known, but a good estimate may
be achieved via Galactic chemical evolution models.
The agreement of chemical abundances derived from early B-type stars and 
the protosolar values indicate that there has been no significant local chemical 
enrichment since the formation of the Sun or that the Sun has been formed in 
another region -- see also c.f. \citet{cp10}.

We leave this discussion for the upcoming work on a larger sample of 
early B-type stars in the solar neighbourood distributed in various OB associations 
and the local field (Nieva \& Przybilla, in prep.)

\subsection{Late-type stars in Ori\,OB1}\label{late-type}

Late-type stars are other abundance indicators than the Sun to compare with our present-day 
abundances derived for the Orion star-forming region. The main differences of these stars with 
our Sun is that they have been indeed formed from the same material than the early B-type stars 
studied here and that they should be younger than the Sun because they belong to a young association. 
There are however a few issues to bear in mind when
evaluating chemical abundances of cooler stars. 
Elemental abundances determined for late-type stars are usually not
absolute values, like in this work, but they are derived relative to a
reference, which in most cases is the Sun.
 At present, atmospheric codes accounting for 3D effects, relevant for stars with convective envelopes, 
or a non-LTE treatment of chemical elements are not implemented for most analyses of late-type stars
other than the Sun.
In spite of these issues, for completeness and to have a general picture of the Orion star-forming region, 
we cite a few recent works on the chemical composition of 
late-type stars in Ori\,OB1 and compare them with our results. 

\citet{Gon08}, \citet{Dor09}, and \citet{Bia10} have recently obtained Fe
abundances\footnote{The original papers provide Fe abundances
on a [Fe/H] scale. We used the solar Fe abundances indicated in each study
to compute the Fe abundances in the same scale used in
this work.} in late-type stars in Ori\,OB1b and the Orion Nebula Cluster 
(ONC = Ori\,Ob1d).
The three studies remark the small scatter in the derived abundances 
within each subgroup.
For Ori\,OB1b, \citeauthor{Gon08} and \citeauthor{Bia10} derive 
7.48\,$\pm$\,0.09 and 7.46\,$\pm$\,0.05 dex, respectively, while \citeauthor{Dor09} find  
hints (only 1 star analysed) for subsolar metallicity. For stars in the ONC, 
\citeauthor{Dor09} obtain 7.49\,$\pm$\,0.07 dex, and \citeauthor{Bia10}, 7.41\,$\pm$\,0.11 dex. 
As argued by \citeauthor{Bia10}, these discrepancies may be due to systematic effects in the analysis. 
For example, it is remarkable that the re-analysis by \citeauthor{Bia10} of some of the stars in the ONC analysed by \citeauthor{Dor09} led to differences up to 0.13 dex in some cases.
Our present-day abundance of iron is in agreement with results
from the analysis of late-type stars in Ori\,OB1.

\section{Summary and conclusions}\label{summary}

 We determined the stellar parameters and H, He, C, N, O, Ne, Mg, Si and Fe abundances 
of a sample of 13 early B-type stars in the Ori\,OB1 star-forming region by means
of a detailed, self-consistent spectroscopic analysis. We applied an updated 
spectral synthesis in non-LTE based on a hybrid non-LTE approach using 
{\sc Atlas9} line blanketed LTE model atmospheres, non-LTE {\sc Detail+Surface} line 
formation calculations, and line-fitting analysis techniques. We used lines from different 
ionization stages of these elements where possible, and treated carefully several sources 
of systematic errors -- summarized in \citet{np10a, np10b} -- in order to obtain accurate and reliable
results and minimize as much as possible the final uncertainties.

The stellar parameters derived here agree with those determined 
in Paper~I of this series using the non-LTE, line-blanketed stellar atmosphere and wind code 
{\sc Fastwind}. The low dispersion of O and Si abundances in B-type stars in the Orion OB1 
association found in Paper~I is confirmed. This is the first systematic comparison of results 
derived from the codes {\sc Fastwind} and {\sc Atlas+Detail+Surface} -- commonly used in the 
analysis of massive stars -- from a high quality set of stellar spectra. Moreover, this work 
constitutes the first step for further comparisons of elements to be incorporated in {\sc Fastwind}, 
that is also able to analyse more massive stars with winds.

The present work finds a significantly smaller abundance scatter (mainly in C, N, O, and Si, but also
Fe) in Ori\,OB1 early B-type stars than previous works and similar scatter than CHL06 for Ne. The derived metal abundances 
are in excellent agreement with results of early B-type stars in the solar neighbourhood that define the
Present-Day Cosmic Abundance Standard (PNB08). They also agree with data from an extended sample 
of stars in the solar vicinity (Nieva \& Przybilla, in prep.). Despite the different state of these massive young 
stars and of older stars like the Sun in terms of Galactic chemical evolution, good agreement with the 
solar values is found -- which poses the question of the origin of the Sun. Late-type stars in the region 
also present similar iron abundances.

This study, together with Paper~I, has allowed us to establish a precise, homogeneous and reliable set of abundances 
for C, N, O, Ne, Mg, Si and Fe in the early B-type stars of Orion OB1.  In a first application, the present results have facilitated constraints 
on the dust-phase composition of the Orion \hii\ region derived in Paper~II.

\begin{acknowledgements}
We would like to thank N. Przybilla for providing us with his updated versions of model atoms, valuable comments and careful reading of the manuscript. We also acknowledge M. Asplund, G. Stasi\'nska, J. Puls and F. Najarro for their suggestions to improve the manuscript.
\end{acknowledgements}

%%%%%%%%%%%%%%%%%%%%%%%%%%%%%%%%%%%%%%%%%%%%%%%%%%%%%%%%%%%%%%%%%%%%%%%%%%%%%%%%%
%%%%%%%%%%%%%%%%%%%%%%%%%%%%%%%%%%%%%%%%%%%%%%%%%%%%%%%%%%%%%%%%%%%%%%%%%%%%%%%%%

\Online

\appendix
\section{An example of a global fit: HD\,35299}\label{appA}

To illustrate the global quality achieved in the spectral synthesis
when a final self-consistent solution is considered, we show in
Figs.~\ref{39_45}-\ref{63_69} the observed spectrum of an example star, HD\,35299 and a 
synthetic spectrum computed using the stellar parameters and abundances indicated 
in Table \ref{parameters}. 
The spectral analysis was based on all Balmer and He lines and 
the metal lines marked at the bottom of the panels.
For completeness purposes, all available lines from our present model atoms  -- even the lines excluded from the analysis for different reasons mentioned in Sect.~\ref{analysis} -- and also from other atoms not analysed here in NLTE are included. 

Surface gravities have been determined from the wings of the Balmer lines and via multiple 
metal ionization equilibria, that can be even more sensitive to this parameter. The cores 
of the hydrogen lines --  which are not always well fitted -- do not affect the analysis 
performed in this work, or at least, the effect is negligible.

The line identification is based on the lines included in our input for {\sc Surface}. 
Evidently, some remaining lines/elements still need to be included to achieve complete 
coverage by the global spectrum synthesis. Identifications of the missing lines in the 
most important blue spectral region may be achieved on the basis of the spectral atlases
of \citet{kilian91}.

\setcounter{figure}{2}

\begin{figure*}[h!]
\centering
\resizebox{0.85\hsize}{!}{\includegraphics[angle=0]{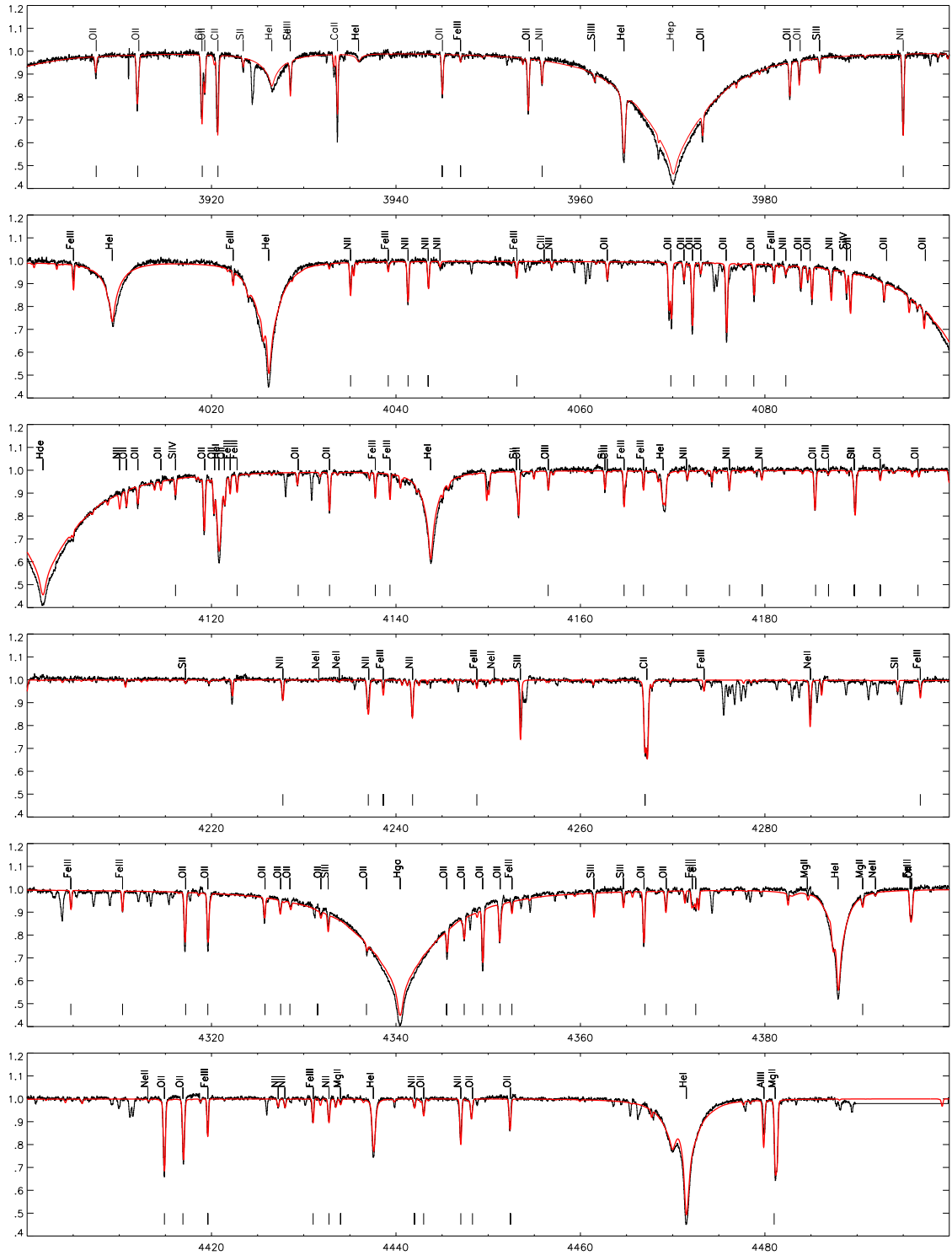}}\\
\caption[]{
Comparison between the observed (black) and the model spectrum (red line) 
for the star HD\,35299, for parameters and elemental abundances as
given in Table 3. The spectral analysis was based on 
all Balmer and He lines and the metal lines marked at the bottom of the panels. Additional lines have been included 
in the global spectrum for visualization purposes, but the linelist is
not complete. See the text for details.}
\label{39_45}
\end{figure*}

\begin{figure*}
\centering
\resizebox{0.89\hsize}{!}{\includegraphics[angle=0]{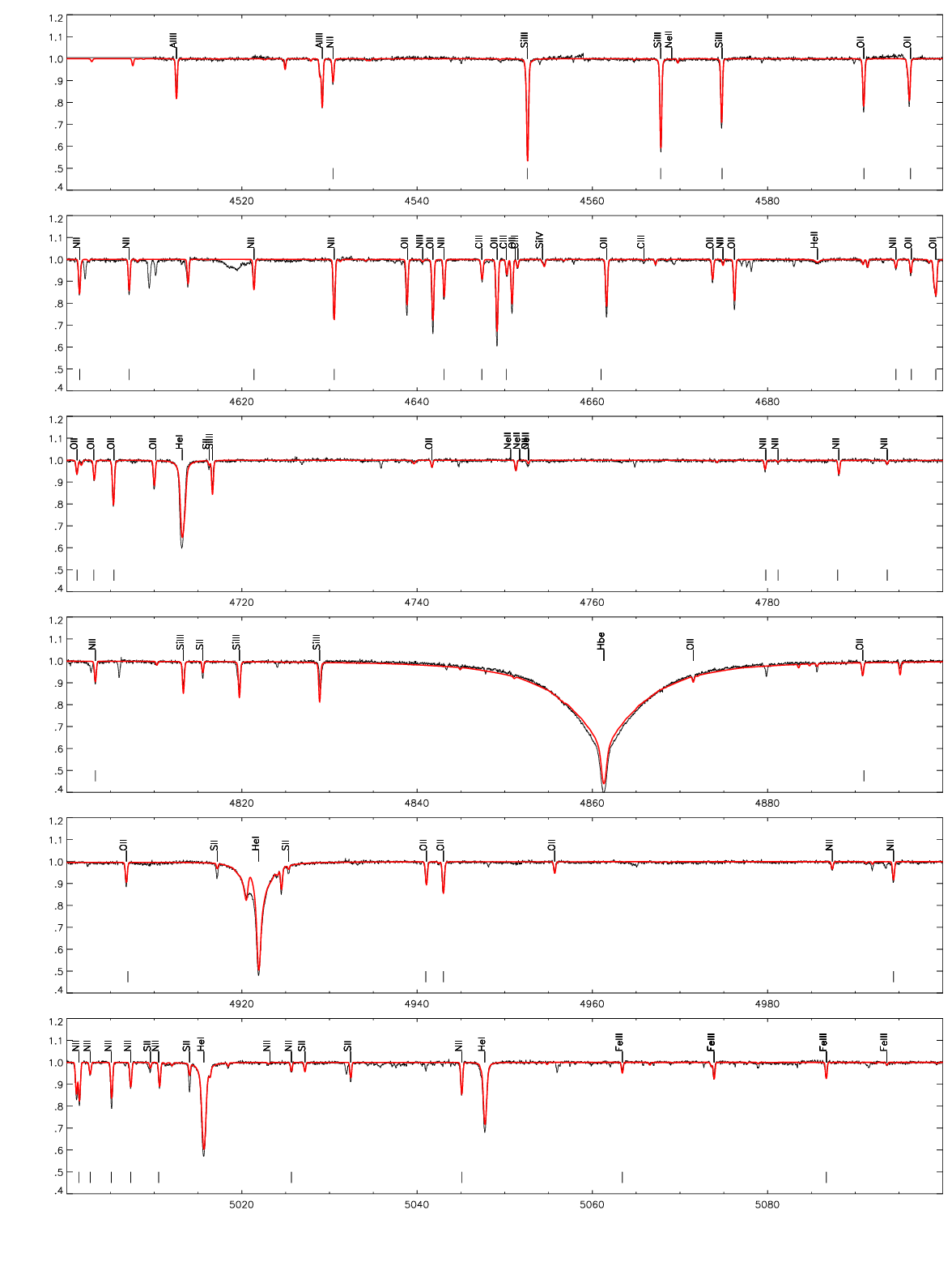}}\\
\caption[]{Same as Fig.~\ref{39_45}} 
\label{45_51}
\end{figure*}

\begin{figure*}
\centering
\resizebox{0.89\hsize}{!}{\includegraphics[angle=0]{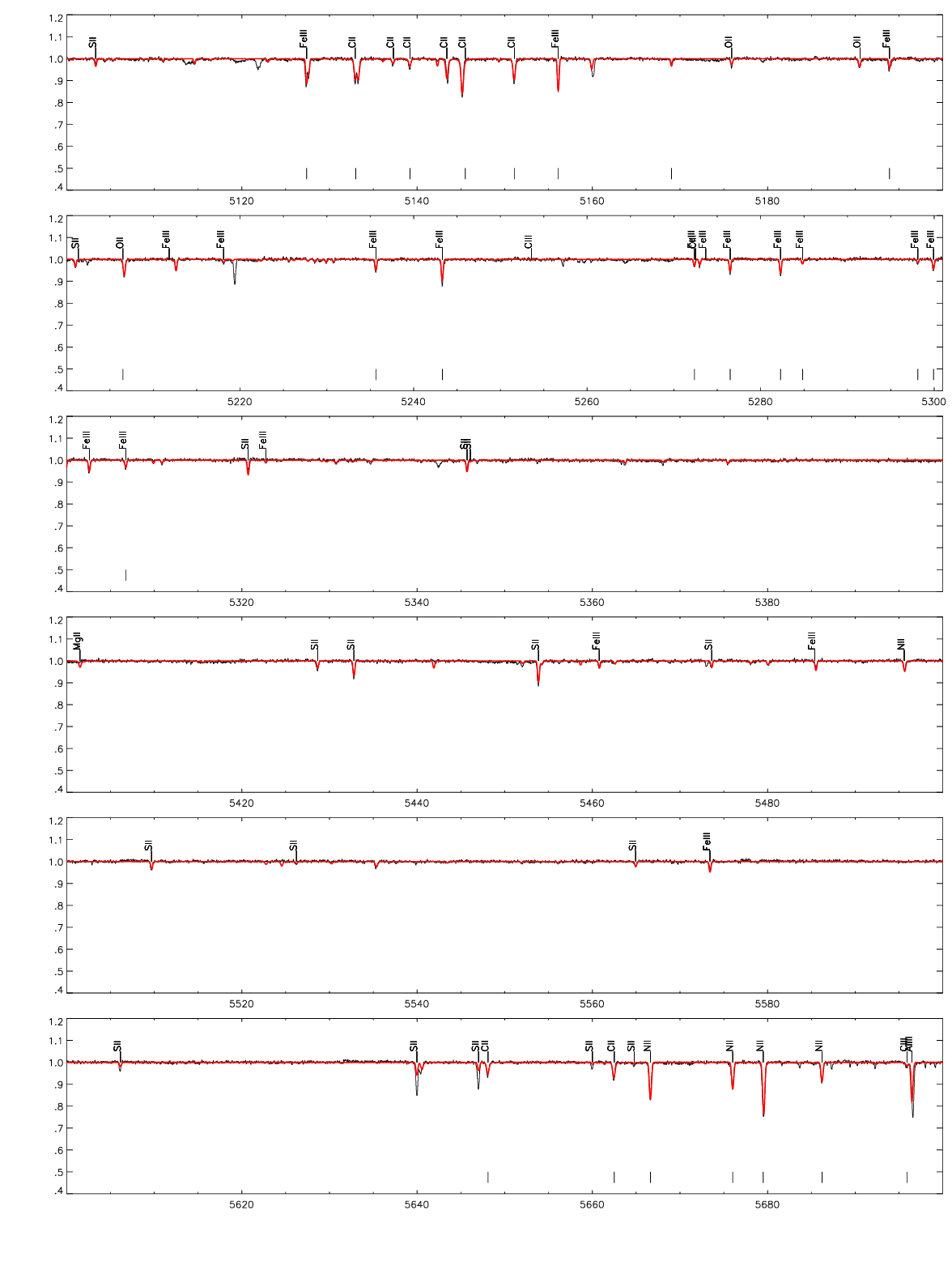}}\\
\caption[]{Same as Fig.~\ref{39_45}.} 
\label{51_57}
\end{figure*}

\begin{figure*}
\centering
\resizebox{0.89\hsize}{!}{\includegraphics[angle=0]{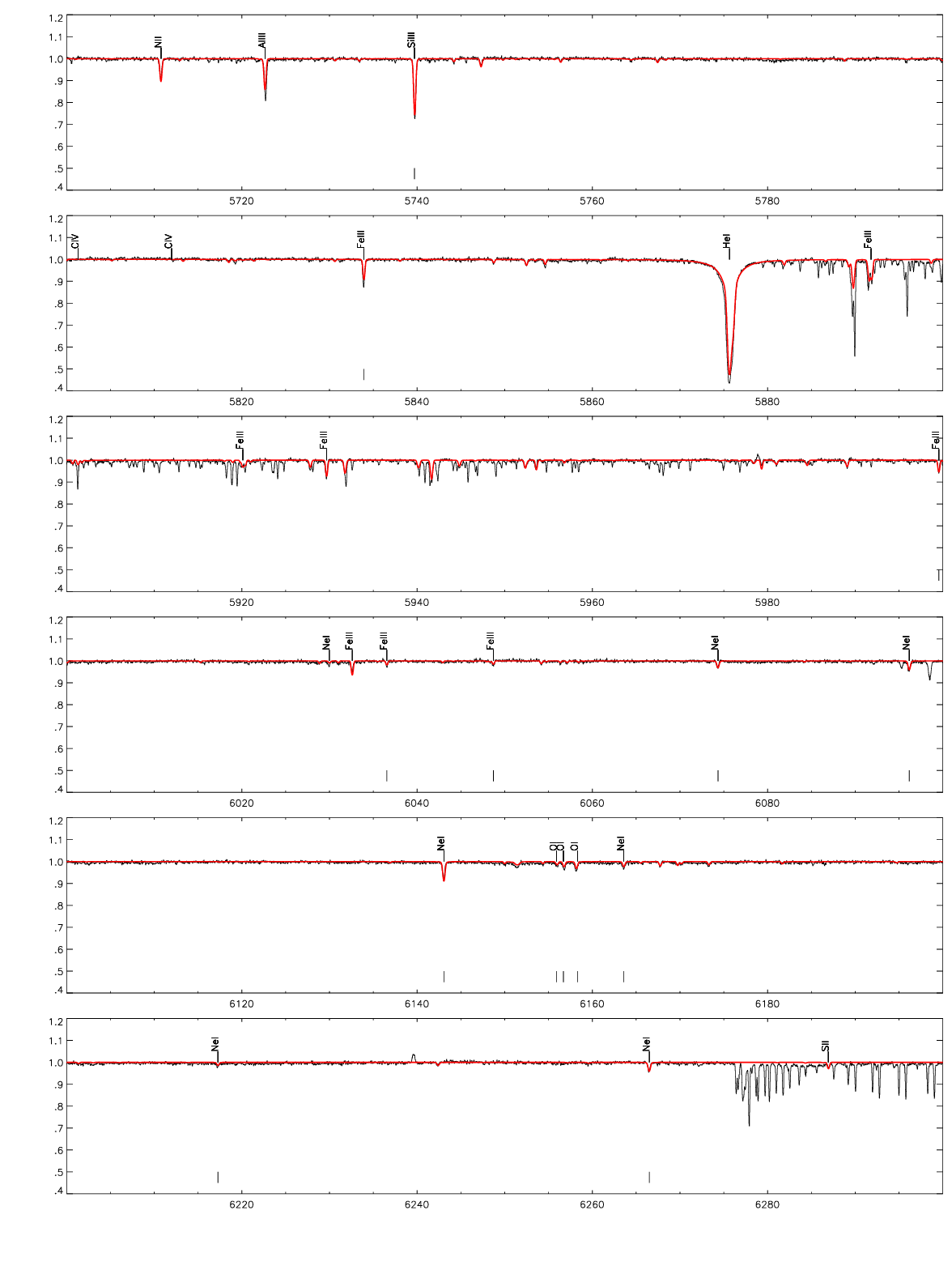}}\\
\caption[]{Same as Fig.~\ref{39_45}.} 
\label{57_63}
\end{figure*}

\begin{figure*}
\centering
\resizebox{0.89\hsize}{!}{\includegraphics[angle=0]{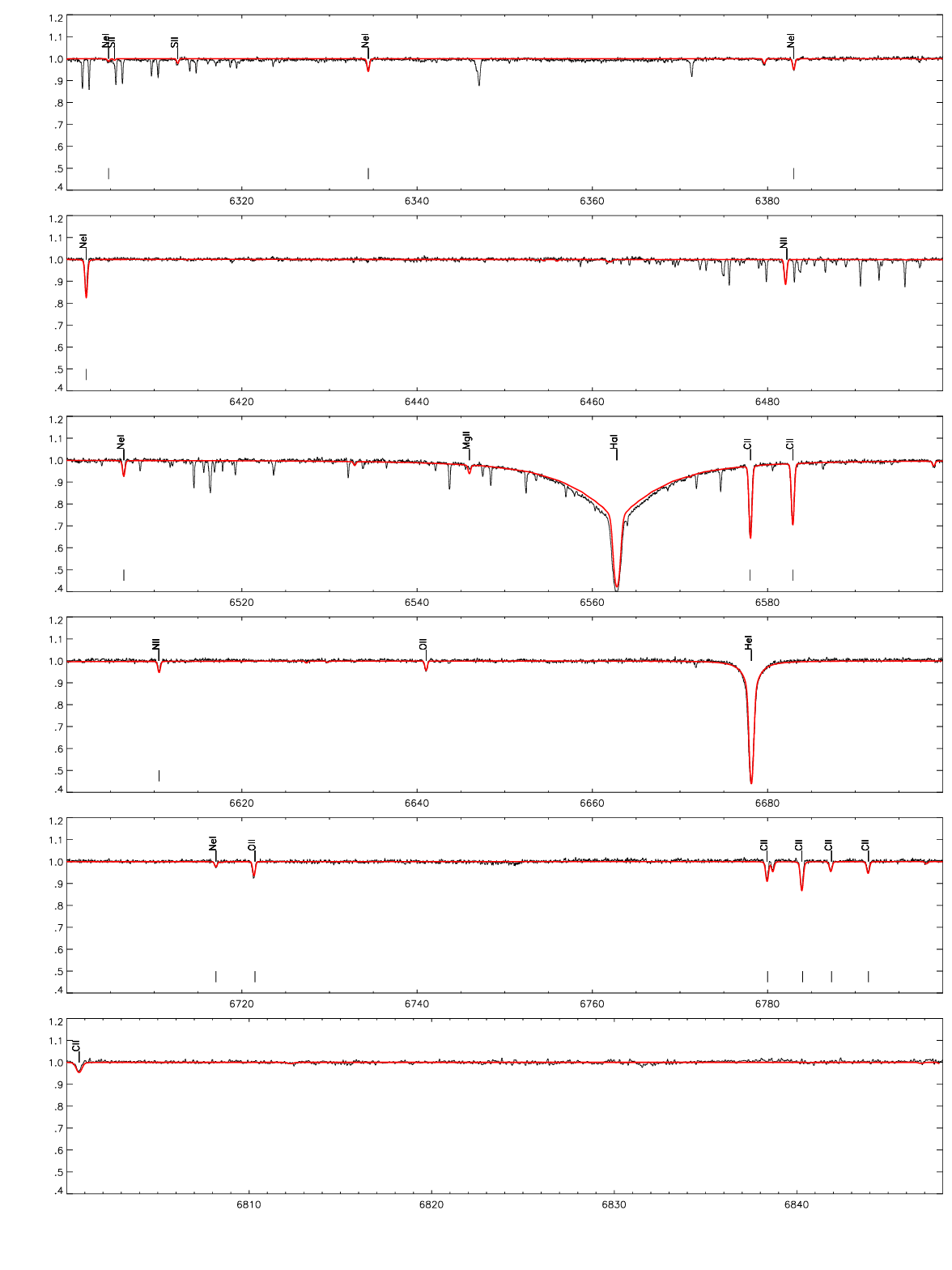}}\\
\caption[]{Same as Fig.~\ref{39_45}.}
\label{63_69}
\end{figure*}

\end{document}